\documentclass[11pt]{article}
\usepackage{graphicx}
\usepackage{amssymb}
\usepackage{amsmath}
\usepackage{authblk}
\usepackage{lineno}
\usepackage{hyperref}
\def\fracd{\displaystyle\frac}
\def\sumd{\displaystyle\sum}

\begin{document}
%\begin{frontmatter}

\title{ Negative Thermal Expansion in Solid Deuteromethane}

\author[$\dagger$]{Yu.A. Freiman}

\author[$\dagger$]{V.V. Vengerovsky}

\author[$\star$]{ Alexander F. Goncharov}

\affil[$\dagger$]{ B.Verkin Institute for Low Temperature Physics and Engineering
of the National Academy of Sciences of Ukraine, 47 Nauky Avenue, Kharkiv 61103, Ukraine}
\affil[$\star$]{Geophysical Laboratory, Carnegie Institution of Washington, 5251 Broad Branch Road NW, Washington D.C. 20015, USA}
\maketitle
\begin{abstract}
The thermal expansion at constant pressure of solid CD$_4$ III is
calculated for the low temperature region where only the
rotational tunneling modes are essential and the effect of
phonons and librons can be neglected. It is found that in mK
region there is a giant peak of the negative thermal expansion.
The height of this peak is comparable or even exceeds the thermal
expansion of solid N$_2$, CO, O$_2$ or CH$_4$ in their triple
points. It is shown that like in the case of light methane, the
effect of pressure is quite unusual: as evidenced from the
pressure dependence of the thermodynamic Gr\"{u}neisen parameter
(which is negative and large in the absolute value), solid CD$_4$
becomes increasingly quantum with rising pressure.

{\it Keywords:solid deuteromethane, negative thermal expansion, Gr\"{u}neisen thermodynamical parameter}

{\it PACS 65.40.De}
\end{abstract}

%\begin{keyword}

%\PACS 65.40.De
%\end{keyword}

%\end{frontmatter}

%\linenumbers

\section{Introduction}
Heberlain and Adams \cite{Heberlain1970} found that the thermal
expansion coefficient (TEC) at constant pressure, $\beta_P$, of
solid methane CH$_4$, becomes negative below 8.7 K. It should be
noted that Manzhelii et al. \cite{Manzhelii1969} also observed
this effect but ascribed it to problems with thermometry.
Subsequent measurements
\cite{Aleksandrovskii1976,Aleksandrovskii1978} showed that as the
temperature is lowered the absolute value of $|\beta_P|$ continues
to grow and at the lowest temperature achieved in the dilatometric
measurements (2 K) it is still far from a maximum. Negative
thermal expansion in solid methane occurs in the temperature and
pressure ranges, where the contribution of phonons and librons to
thermodynamics can be neglected compared to that of rotational
tunneling states. Due to this tunneling between different
orientations the librational ground state is splitted. In the case
of CH$_4$ this tunneling splittings is about 1 K. The existence of
these splittings has been confirmed in the NMR experiment by
Glattli et al. \cite{Glattli1972} and by Press and Kolmar
\cite{Press1975} in a neutron scattering experiment. Transitions
between different states simultaneously include changes in spin
and orientation states.

It is easy to see why the rotational tunneling gives rise to the
negative thermal expansion. Indeed, the volume change of the
crystal with changing temperature (in the temperature range where
the  contribution of phonons is negligible) is determined by
the competition between two factors. The contribution to the free
energy due to populating of the rotational tunnel states of the
ordered sublattices  on rising temperature favors a contracting of
the lattice. The height of the potential barriers separating the
equivalent minima then increases, the magnitude of the tunneling
splitting decreases and the crystal free energy decreases. This
effect is counterbalanced by the loss in elastic energy increasing
with increased contraction, which stabilizes the crystal volume at
each temperature. Thus, this mechanism of the negative thermal
expansion is purely quantum.

The theoretical maximum of $|\beta_P|$ at zero pressure which lays
at about 0.66 K is about 1.1 $\times 10^{-3}$ {\rm K}$^{-1}$
\cite{Freiman1983}. The effect of the hydrostatic pressure on the
thermal expansion of solid CH$_4$ was calculated in Ref.
\cite{Freiman2020}. It was shown that the magnitude of the peaks
of the negative thermal expansion increases with pressure while
the peaks shift to the low-temperature region. The effect of
pressure can be seen from the following figures: at 1.9 kbar the
position of the peak lays at 0.169 K while the height of the peak
is 1.77 $\times 10^{-3}$ {\rm K}$^{-1}$. It should be noted that
it is quite unusual that the quantum effects increase with
pressure.

Both CH$_4$ and CD$_4$ have three long-lived spin species, $A$,
$T$, and $E$ which transform under the group $T$ as  one $A$
representation, three $T$ representations, and one two-dimensional
$E$ representation. The experimental observations of the tunneling
states in phase II of CH$_4$ (73 and 143 $\mu$eV) \cite{Press1975}
have been used to predict the effect on tunneling of an isotopic
shift of CH$_4$ to CD$_4$. A conclusion reached was that the
tunnel splittings in CD$_4$ would be reduced by about a factor of
50 from those in CH$_4$. Numerous studies are devoted to the
investigations of the effects which have place when instead of
solid light methane CH$_4$  heavy solid methane CD$_4$ was
considered \cite{Prager1997}.

When nuclear spin and the requirements of the permutation symmetry
are taken into account, it is found that only certain values of
the total nuclear  spin $I$ are associated with each spin species.
For CD$_4$, the resulting degeneracies of the tunneling states
(and associated $I$ values) of the three  spin species are as
follows: $A$, 15 ($I$ =0, 2, 4); $T$, 54 ($I$ =1, 2, 3); $E$, 12
($I$ =0, 2).

The $P-T$ phase diagram of solid  deuteromethane CD$_4$ was
studied by van der Putten et al. \cite{VanderPutten1984}  by using
the nuclear magnetic resonance at pressures up to 8 kbar and
temperatures 10 - 70 K. The phase diagrams of CH$_4$ and CD$_4$
are qualitatively similar \cite{Hebert1987}, except that the phase
III of CD$_4$ is stable down to zero pressure while phase III in
CH$_4$ exists only under pressure \cite{Shubnikov1939}.  When
temperature decreases, at 22.15 K CD$_4$ shows a further
transition to a slightly tetragonally distorted phase
\cite{Collwell1963}, whereas CH$_4$ remains in phase II down to
the lowest temperatures. Such difference in the phase diagrams of
light and heavy methanes results from the quantum effect of the
molecular rotation, since the rotational constant $B_{\rm rot} =
\hbar^2/2I$ of CH$_4$ is twice as large as that of CD$_4$ (5.3 and
2.6 cm$^{-1}$, respectively). The structure of the tetragonal
phase in CD$_4$ was proposed in the x-ray study by Prokhvatilov
and Isakina \cite{Prokhvatilov1979}. The dilatometry studies of
the lattice parameters, molar volume, and thermal expansion of
solid CD$_4$ down to 2 K were performed  by Manzhelii group
\cite{Manzhelii1969,Tolkachev1977}. The x-ray low-temperature data
down to 4.4  K were obtained by Baer et al. \cite{Baer1978}. The
low-temperature measurements of the heat capacity of solid CD$_4$
were carried out by Colwell, Gill, and Morrison. (0.28 - 4.0 K)
\cite{Collwell1962,Collwell1963}, Colwell \cite{Collwell1969},
White and Morrison (0.15 - 4.0 K) \cite{White1980}, and White
\cite{White1982}.

\section{Negative thermal expansion}

In the two-site approximation free energy associated with the
rotational tunneling modes can be written as a sum of the
rotational free energy and the elastic energy:
\begin{equation}\label{free_energy}
{\cal F}(V,T) =  -\frac{1}{2}N k_BT\sumd_{\varepsilon\in \{2,m\}}\ln{Z}_{\varepsilon} +
\frac{(V-V_0)^2}{2\chi V_0},
\end{equation}
where $V_0$ is the volume of the system  at zero temperature and
pressure, $\chi$ is the isothermal compressibility. The summation
in Eq. $\ref{free_energy}$ is taken over the sites belonging to
symmetries 2 and $m$. The statistical sum $Z_{\varepsilon}$ can be
obtained by the summation over the spectrum of the tunneling
rotational modes:
\begin{equation}\label{spectrum}
{Z}_{\varepsilon}=\sumd_{i=0}^4 \alpha_i \exp
\left(\frac{-c^{(\varepsilon)}_i\Delta(V)}{{k_BT}}\right),
\end{equation}

where $\alpha_i$ is the degeneracy of the respective rotational
state, $\Delta(V)$ is a difference of the arithmetic mean  of the
$T$ state energies and the  $A$ state energy, and the constants
$\{c^{(\varepsilon)}_i\}_{i=0}^{4}$ can be obtained from the
following linear equations:
$\Delta(V)=1/3(E_{T_1}+E_{T_2}+E_{T_3})-E_A$,
$E^{(\varepsilon)}_A-E^{(\varepsilon)}_A=c^{(\varepsilon)}_0\Delta(V)$,
$E^{(\varepsilon)}_{T_1}-E^{(\varepsilon)}_A=c^{(\varepsilon)}_1\Delta(V)$,
$E^{(\varepsilon)}_{T_2}-E^{(\varepsilon)}_A=c^{(\varepsilon)}_2\Delta(V)$,
$E^{(\varepsilon)}_{T_3}-E^{(\varepsilon)}_A=c^{(\varepsilon)}_3\Delta(V)$,
$E^{(\varepsilon)}_{E}-E^{(\varepsilon)}_A=c^{(\varepsilon)}_3\Delta(V)$.

The contribution of the tunneling states into pressure is given
by the equation
\begin{equation}\label{defP}
 P= -\left(\frac{\partial{\cal F}}{\partial
V}\right)_T.
\end{equation}

From Eq. (\ref{defP}) we have the following equation:
\begin{equation}\label{P}
P = -\fracd{N}{2}\cdot \frac{\partial \Delta}{\partial V}
\sumd_{\varepsilon\in \{2,m\}} \fracd{{ Y}_{\varepsilon}}{{
Z}_{\varepsilon}} -\frac{V-V_0}{\chi V_0},
\end{equation}
where
\begin{equation}
{ Y}_{\varepsilon}=\sumd_{i=0}^4
\alpha_i c^{(\varepsilon)}_i \exp \left(\frac{-c^{(\varepsilon)}_i\Delta(V)}{{k_BT}}\right).
\end{equation}
The coefficient of thermal expansion at constant pressure
\begin{equation} \beta_P = \frac{1}{V}\left(\frac{\partial
V}{\partial T}\right)_P.
\end{equation}

Let us turn from the variables  $V, T$ to the variables $P, V$
using the Jacobian of the transformation $D(P, V)/D(T, V)$. As a
result, we have
\begin{equation}
\beta_P = -\left(\frac{\partial P}{\partial
T}\right)_V\,\frac{1}{V(\partial P/\partial V)}_T.
\end{equation}

From Eq. (\ref{P}) we have the following relations for the
derivatives $(\partial P/ \partial T)_V$ and $(\partial
P)/\partial V)_T$:
\begin{equation}\label{dPdT}
\frac{\partial P}{\partial T}=-\fracd{N\Delta (\partial \Delta / \partial V)}{2k_B T^2}\sumd_{\varepsilon\in \{2,m\}}\left(\fracd{X_{\varepsilon}}{Z_{\varepsilon}}-\fracd{Y^2_{\varepsilon}}{Z^2_{\varepsilon}}\right)
\end{equation}
\begin{equation}\label{dPdV}
\frac{\partial P}{\partial V}=-\frac{1}{\chi V_0}-\fracd{N}{2}\sumd_{\varepsilon\in \{2,m\}}\left(\fracd{Y_{\varepsilon}}{Z_{\varepsilon}}\frac{\partial^2 \Delta}{\partial V^2}-\frac{1}{k_B T}\left\{\fracd{X_{\varepsilon}}{Z_{\varepsilon}}-\fracd{Y^2_{\varepsilon}}{Z^2_{\varepsilon}}\right\}\left(\frac{\partial \Delta}{\partial V}\right)^2\right)
\end{equation}
\begin{equation}
{X}_{\varepsilon}=\sumd_{i=0}^4 \alpha_i
\left(c^{(\varepsilon)}_i\right)^2 \exp
\left(\frac{-c^{(\varepsilon)}_i\Delta(V)}{{k_BT}}\right).
\end{equation}

Finally,  for the coefficient of thermal expansion we have the
following equation:
\begin{equation*}
\beta_P = -\fracd{N \chi V_0\Delta (\partial \Delta / \partial V)}{2k_B T^2 V}\sumd_{\varepsilon\in
\{2,m\}}\left(\fracd{X_{\varepsilon}}{Z_{\varepsilon}}-\fracd{Y^2_{\varepsilon}}{Z^2_{\varepsilon}}\right)\times
\end{equation*}
\begin{equation}\label{beta}
 \left\{1+\fracd{N\chi V_0}{2}\sumd_{\varepsilon\in
\{2,m\}}\left(\fracd{Y_{\varepsilon}}{Z_{\varepsilon}}\frac{\partial^2
\Delta}{\partial V^2}-\frac{1}{k_B
T}\left\{\fracd{X_{\varepsilon}}{Z_{\varepsilon}}-\fracd{Y^2_{\varepsilon}}{Z^2_{\varepsilon}}\right\}\left(\frac{\partial
\Delta}{\partial V}\right)^2\right)\right\}^{-1}.
\end{equation}

Equation (\ref{beta}) should be supplemented by equations for
$\Delta$ and $\partial \Delta/\partial V$. We shall use for the
further calculations the dependence of the energy  of the
tunneling state $\Delta$ at zero pressure on the rotational
barrier hight $U$, obtained  by H\"{u}ller and Raich
\cite{Huller1979}. Taking into account that the repulsive forces
make the  largest contribution to the derivative $\partial
\Delta/
\partial V$ and assuming that there is a power law relation $U
\sim r^{-n}$ and taking $n=15$ \cite{Nijman1977} we finally have:

\begin{equation}\label{Delta(v)}
\Delta(V) = \omega_0 e^{-\gamma U_0 \left(V/V_0\right)^{-5}},
\end{equation}
where $\omega_0=\Delta(V_0)e^{\gamma U_0}$. Here $U_0$ is the reduced value of the barrier at zero pressure
and temperature.

From Eq. \ref{Delta(v)} we have the following equations:
\begin{multline}
(\partial \Delta /\partial V)_{P,T} = 5 (\Delta / V_0)(\gamma
U_0)\left(\frac{V_0}{V}\right)^6, \\  (\partial^2\Delta/\partial V^2)_{P,T} = 25
(\Delta/V_0^2)(\gamma U_0)^2\left(\frac{V_0}{V}\right)^{12}-30
(\Delta/V_0^2)(\gamma U_0)\left(\frac{V_0}{V}\right)^{7}.
\end{multline}

The effect of pressure on the thermal expansion of solid
deuteromethane can be seen from Fig. \ref{fig: Pressure effect on
thermal expansion}.
\begin{figure}
    \begin{center}
        \includegraphics[scale=0.5]{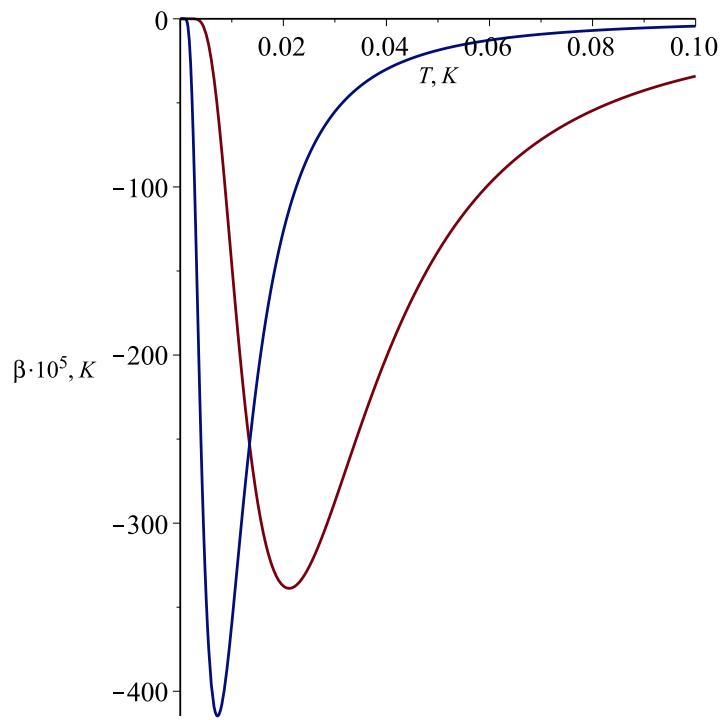}
    \end{center}
    \caption{The effect of pressure on the thermal expansion of solid deuteromethane.
The volume expansion coefficient as a function of temperature: red
curve - zero pressure;
   blue curve - 840 bar.}
    \label{fig: Pressure effect on thermal expansion}
    \end{figure}

The sensitivity of the respective frequency spectrum to the
lattice expansion is described by the Gr\"{u}neisen parameter G
\begin{equation}
G = \beta_P V/C_V\chi,
\end{equation}
where $C_V$ describes the contribution of the respective modes to
the heat capacity $C_V= -T(\partial^2/\partial T^2)$. For the
rotational tunneling modes from  Eq. (\ref{free_energy}) we have
\begin{equation}
C_V^{\rm rot} =
\fracd{N  \Delta^2 }{2k_B T^2 }\sumd_{\varepsilon\in \{2,m\}}\left(\fracd{X_{\varepsilon}}{Z_{\varepsilon}}-\fracd{Y^2_{\varepsilon}}{Z^2_{\varepsilon}}\right)
\end{equation}

Finally, we have
\begin{equation}
G(P)= -\frac{V_0}{\Delta}\frac{\partial \Delta}{\partial V}
 \cdot \left\{1+\fracd{N\chi V_0}{2}\sumd_{\varepsilon\in \{2,m\}}\left(\fracd{Y_{\varepsilon}}{Z_{\varepsilon}}\frac{\partial^2 \Delta}{\partial V^2}-\frac{1}{k_B T}\left\{\fracd{X_{\varepsilon}}{Z_{\varepsilon}}-\fracd{Y^2_{\varepsilon}}{Z^2_{\varepsilon}}\right\}\left(\frac{\partial \Delta}{\partial V}\right)^2\right)\right\}^{-1}.
\end{equation}
Taking into account Eq. (\ref{P}), we have
\begin{equation}
V(T=0, P) = V_0 (1-P\chi).
\end{equation}

At zero temperature
\begin{equation}
G(T=0, P) = - \frac{V_0}{\Delta}\left(\frac{\partial \Delta}{\partial
V}\right)_{P=0} = -5\gamma U_0 \left(1-P\chi\right)^{-6} .
\end{equation}
\begin{equation}
G(T=0, P=0) = -34.4.
\end{equation}

\begin{figure}
    \begin{center}
        \includegraphics[scale=0.5]{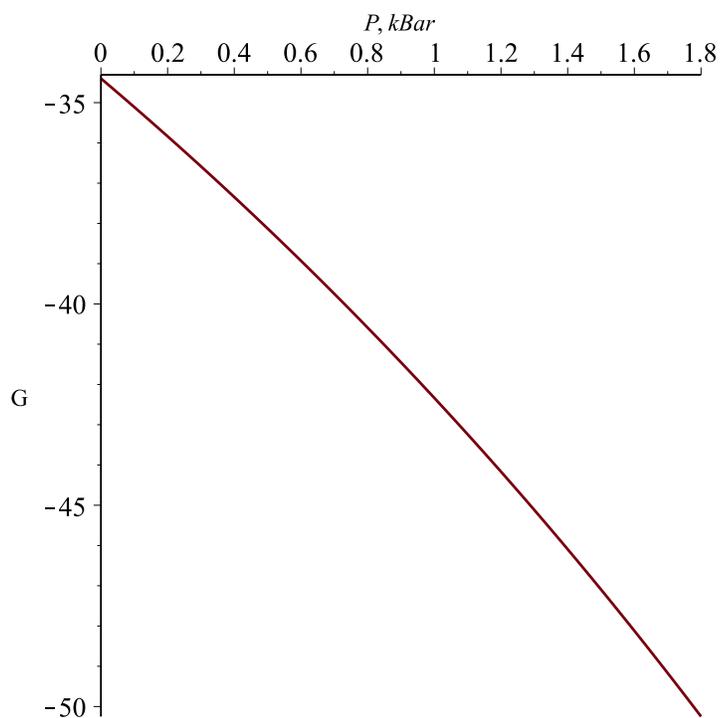}
    \end{center}
    \caption{Thermodynamical Gr\"{u}neisen parameter of solid methane as a function of
    pressure.}
    \label{fig: Gruneisen parameter}
    \end{figure}

The pressure dependence of the thermodynamical
Gr\"{u}neisen parameter $G(P)$ is shown in Fig. \ref{fig: Gruneisen parameter}.

     \begin{figure}\begin{center}
    \includegraphics[scale=0.5]{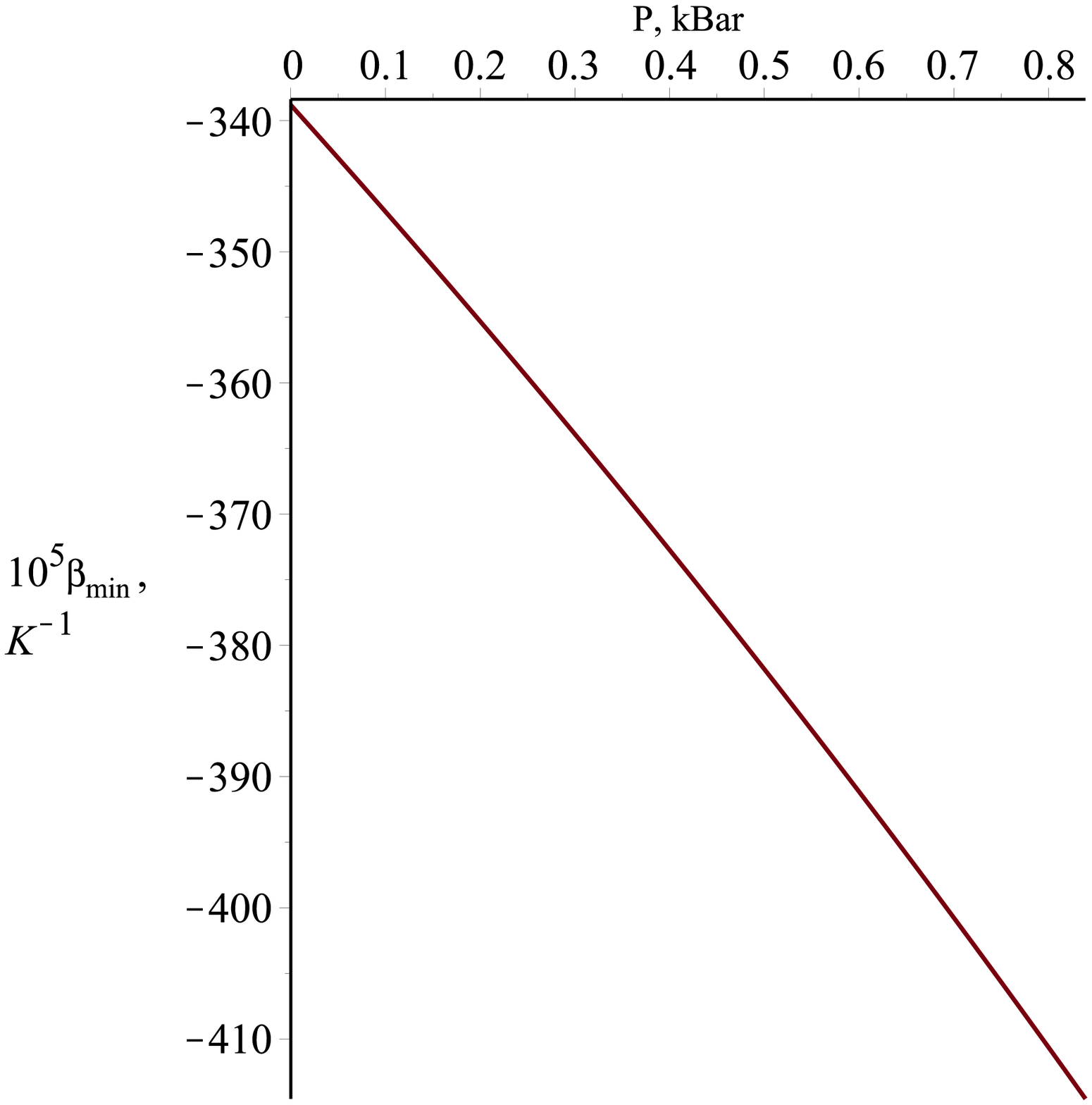}  \hfill
    \caption{  The minimal volume expansion coefficient  as a function of pressure.} \label{fig: minimum_beta}
    \end{center}
    \end{figure}
    \begin{figure}
    \begin{center}
        \includegraphics[scale=0.5]{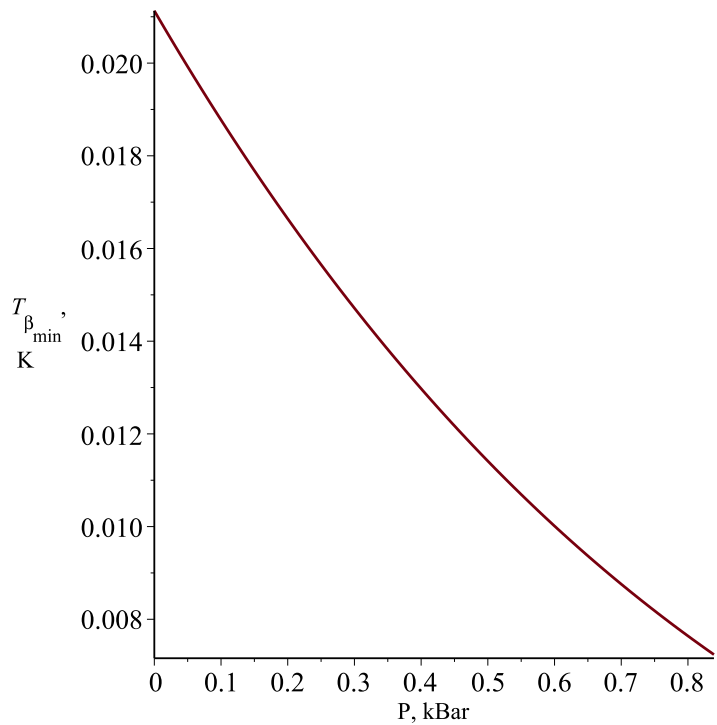}
    \end{center}
          \caption{The temperature of the minimal volume expansion coefficient  as a function of pressure.}  \label{fig: Location_of_minimum_beta}
    \end{figure}

\section{Numerical results}

The rotational tunneling in phase III of CD$_4$ has been studied
by the incoherent neutron scattering in numerous papers
\cite{Press1980,Prager1981,Buchman1982,Fabre1982,Thiery1985,Calvani1989,
Calvani1989A,Prager1990,Prager1997,Prager2002,Huller2008}. A
rather complicated structure of the inelastic neutron scattering
spectrum in CD$_4$ compared to that in CH$_4$ can only be
explained assuming that a superposition of the several single
molecule spectra takes place. Prager, Press, and Heidemann
\cite{Prager1981} proposed a number of the alternative models of
the tunneling splittings present in CD$_4$ III, based on
the incoherent neutron scattering data. With the use of the
notation of H\"{u}ller  \cite{Huller1977}, a given orientational
potential is characterized by four overlap matrix elements $h_i$,
$h_1$ $h_2$, $h_3$, $h_4$ for 120$^ô$ rotations about the four
threefold axes of the molecule, and by three overlap matrix
elements $H_x, H_y, H_z$, for 180$^o$ rotations about the three
twofold axes of the molecule. In all of these models, the 180$^o$
overlap matrix elements are considered negligible compared to the
120$^o$ overlap matrix elements. Under this assumption, the energy
of the $E$ states is $-4h = -(h_1+h_2+h_3+h_4)$, the energy of the
$A$ states is 8$h$, and the mean energy of the $T$ states is zero.
All the overlap matrix elements are negative, so the $A$ states
are lowest in energy.

In our numerical calculations, we used the two-site model developed
by Prager et al. \cite{Prager1981} for  symmetries 2 and $m$.
For the strength parameters $h_i$ we used the values obtained from
their fit to the inelastic scattering data (see Table II
\cite{Huller2008}).

The following values of the parameters were used in the
calculations:  $\chi = 3.4\cdot 10^{-11}cm^2/{dyn}$, $\gamma
U_0=6.88$, twice as large as in light methane \cite{Freiman1983}
The minimal value of the thermal expansion of solid deuteromethane
as a function of pressure can be seen in Fig. \ref{fig:
minimum_beta}. The temperature of the minimum  of the thermal
expansion of solid deuteromethane as a function of pressure can be
seen in Fig. \ref{fig: Location_of_minimum_beta}.

Concerning the comparison with experiment, the situation is as
follows.  In the paper \cite{Freiman1983} the author compared the
proposed theory with the available at that time experimental data
for light methane. Already at that moment, the experiment has been
falling behind the theory. Moreover, this is true today. With
regard to deuteromethane, any exparimental data on the thermal
expansion in the range below 2 K are absent. Meanwhile, the giant
peak of the negative thermal expansion predicted in our article
and the pressure effects are quite accessible to modern
low-temperature experiment.

\section{Conclusions}

The thermal expansion at constant pressure of solid deuteromethane
in the low-temperature phase CD$_4$ III is calculated for the low-temperature region where the contributions from phonons and
librons can be neglected and the whole effect of the thermal
expansion is due to the rotational tunneling modes. It is found
that in the mK region there is a giant peak of the negative
thermal expansion. The height of this peak is so large that it is
comparable or even exceeds the thermal expansion of the simple
molecular solids N$_2$, CO, O$_2$, or CH$_4$ in their triple points
\cite{Manzhelii1996, Manzhelii1998}. The effect of pressure is
calculated and it is shown that like solid CH$_4$ solid CD$_4$
becomes increasingly quantum with rising pressure. Such unusual
behavior is due to the fact that the rotational tunneling modes
go down with rising pressure.

\end{document}